\documentclass[aps, prd, amsmath, floats, floatfix, twocolumn, nofootinbib, superscriptaddress]{revtex4-1}
\usepackage[pdftex]{graphicx}
\usepackage{color}
\usepackage{latexsym}
\usepackage[colorlinks]{hyperref}
\usepackage{amsthm,amsmath,amssymb}
\usepackage{mathrsfs}
\usepackage{makecell}

\begin{document}

\title{Impact of the returning radiation in current tests of the Kerr black hole hypothesis using X-ray reflection spectroscopy}

\author{Shafqat~Riaz}
\affiliation{Center for Field Theory and Particle Physics and Department of Physics, Fudan University, 200438 Shanghai, China}
\affiliation{Theoretical Astrophysics, Eberhard-Karls Universit\"at T\"ubingen, D-72076 T\"ubingen, Germany}

\author{Temurbek Mirzaev}
\affiliation{Center for Field Theory and Particle Physics and Department of Physics, Fudan University, 200438 Shanghai, China}

\author{Askar~B.~Abdikamalov}
\affiliation{School of Mathematics and Natural Sciences, New Uzbekistan University, Tashkent 100007, Uzbekistan}
\affiliation{Center for Field Theory and Particle Physics and Department of Physics, Fudan University, 200438 Shanghai, China}
\affiliation{Ulugh Beg Astronomical Institute, Tashkent 100052, Uzbekistan}

\author{Cosimo~Bambi}
\email[Corresponding author: ]{bambi@fudan.edu.cn}
\affiliation{Center for Field Theory and Particle Physics and Department of Physics, Fudan University, 200438 Shanghai, China}

\date{\today} 

\begin{abstract}
The past 10~years have seen remarkable progress in our capability of analyzing reflection features in the X-ray spectra of accreting black holes. Today X-ray reflection spectroscopy is a mature technique and a powerful tool for studying the accretion process around black holes, measuring black hole spins, and testing Einstein's theory of General Relativity in the strong field regime. However, current reflection models still rely on a number of simplifications and caution is necessary when we derive very precise measurements. In this paper, we study the impact of the returning radiation on our capability of measuring the properties of black holes using X-ray reflection spectroscopy, and in particular on our capability of testing the Kerr black hole hypothesis. While the returning radiation alters the reflection spectrum of the disk, from the analysis of our simulations we find that models without returning radiation can normally recover well the correct black hole spin parameters and can test the Kerr metric. 
\end{abstract}


\maketitle

\section{Introduction}

\begin{figure}[b]
\centering
\includegraphics[width=0.45\textwidth,trim=0.0cm 2.5cm 0.0cm 1.0cm,clip]{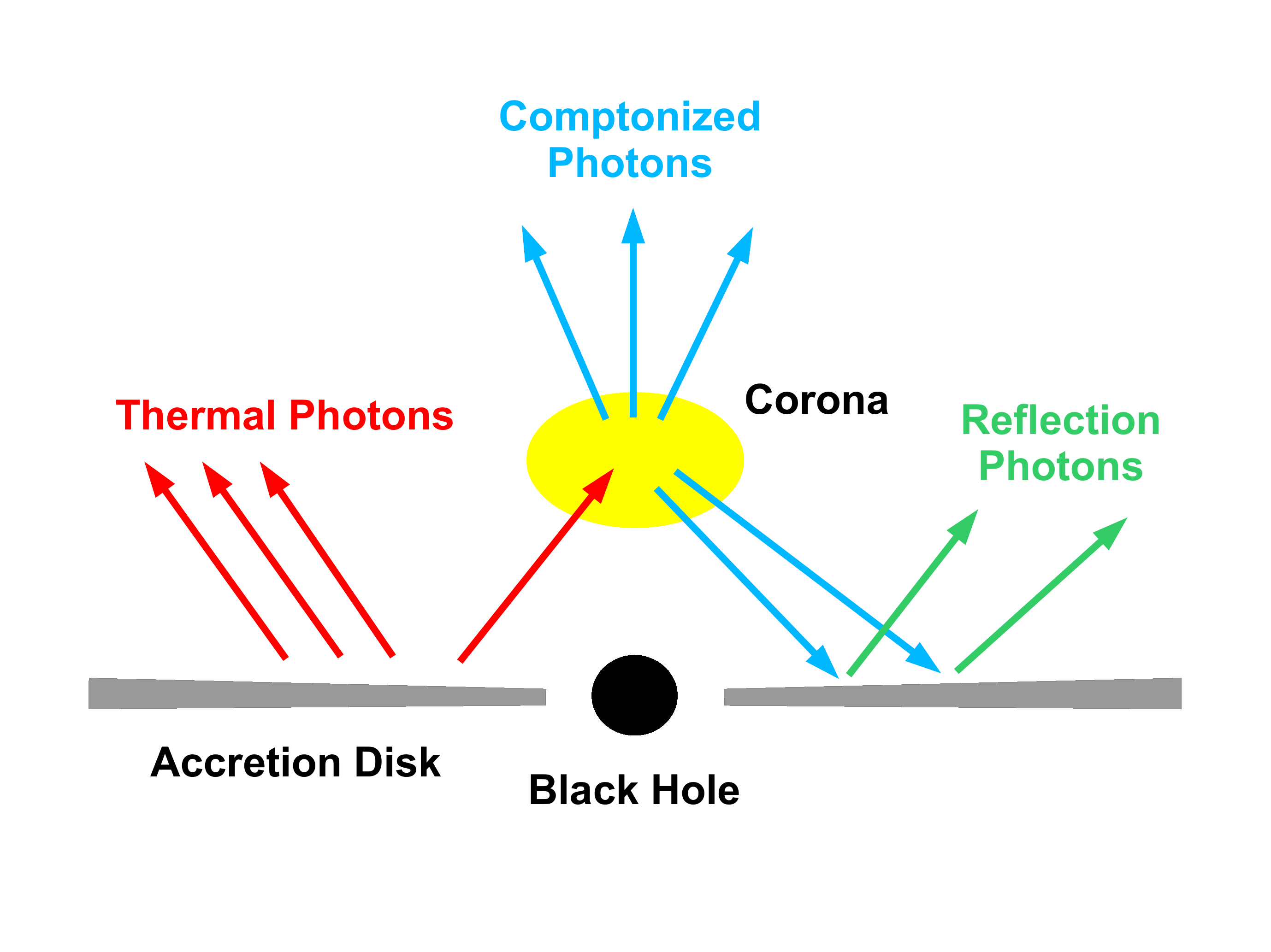}
\caption{Disk-corona system. Figure from Ref.~\cite{Bambi:2021chr} under the terms of the Creative Commons Attribution 4.0 International License. \label{f-corona}}
\end{figure}

Blurred reflection features are common in the X-ray spectra of accreting black holes and are produced by illumination of a ``cold'' accretion disk by a ``hot'' corona~\cite{Fabian:1989ej,Fabian:2000nu,Risaliti:2013cga} (for a recent review, see Ref.~\cite{Bambi:2020jpe}). The prototype of the astrophysical system is shown in Fig.~\ref{f-corona}. A black hole is surrounded by a geometrically thin and optically thick accretion disk. The gas in the disk is in local thermal equilibrium and every point on the surface of the disk has a blackbody spectrum. The whole accretion disk has a multi-temperature blackbody spectrum because the temperature of the gas increases as it falls into the gravitational well of the black hole. The disk is cold because the gas can efficiently cool down by emitting radiation. The spectrum of the disk is normally peaked in the soft X-ray band in the case of stellar-mass black holes in X-ray binary systems and in the UV band in the case of supermassive black holes in active galactic nuclei. The corona is some hotter plasma ($\sim 100$~keV) near the black hole and the central part of the accretion disk, even if its exact morphology is not yet well understood. The corona may be the base of the jet, the hot atmosphere above the accretion disk, the gas in the plunging region between the inner edge of the disk and the black hole, etc. Thermal photons from the disk can inverse Compton scatter off free electrons in the corona. The Comptonized photons can illuminate the disk: Compton scattering and absorption followed by fluorescent emission generate the reflection spectrum.

The reflection spectrum in the rest-frame of the material in the disk is characterized by narrow fluorescent emission lines in the soft X-ray band and a Compton hump with a peak around 20-30~keV~\cite{Ross:2005dm,Garcia:2010iz}. The most prominent emission line is normally the iron K$\alpha$ complex, which is a narrow feature at 6.4~keV for neutral or weakly ionized iron atoms and shifts up to 6.97~keV for H-like iron ions. The reflection spectrum of the whole disk as seen by a distant observer is blurred because it results from the sum of photons emitted from different parts of the accretion disk and affected by different gravitational redshift and Doppler boosting~\cite{Fabian:1989ej,Laor:1991nc,Bambi:2017khi}. X-ray reflection spectroscopy is the analysis of these relativistically blurred reflection features and, in the presence of high-quality data and a sophisticated theoretical model, it is a powerful tool to study the accretion process in the strong gravity region of black holes, measure black hole spins (and it is currently the only robust technique to measure the spins of supermassive black holes), and test General Relativity in the strong field regime~\cite{Bambi:2020jpe}.

In the past 10~years, there have been significant advancements in our capability of analyzing these relativistically blurred reflection features in the X-ray spectra of black holes, thanks to a new generation of theoretical models (e.g., {\tt kyn}~\cite{Dovciak:2003jym}, {\tt reflkerr}~\cite{Niedzwiecki:2007jy,Niedzwiecki:2018wtc}, {\tt relxill}~\cite{Dauser:2013xv,Garcia:2013lxa}, and {\tt reltrans}~\cite{Ingram:2019qlb}) and new X-ray observatories (e.g. \textsl{NuSTAR}~\cite{NuSTAR}). However, all the available reflection models are based on a number of simplifications and, in the presence of high-quality data, such modeling simplifications may lead to systematic uncertainties in the estimate of the model parameters that could exceed their statistical uncertainties. It is thus crucial to understand the impact of these modeling simplifications in the measurements of the properties of black holes. If these modeling uncertainties were not under control, we could have very precise but not very accurate measurements, which could easily lead to incorrect physical interpretations.

In the past few years, our group has developed the model {\tt relxill\_nk}~\cite{Bambi:2016sac,Abdikamalov:2019yrr}, which is an extension of the {\tt relxill} package~\cite{Dauser:2013xv,Garcia:2013lxa} specifically designed to test the Kerr black hole hypothesis (i.e., whether the spacetime geometry around black holes is described by the Kerr solution as expected from General Relativity and in the absence of exotic fields)~\cite{Bambi:2015kza}. {\tt relxill\_nk} has been used to analyze a number of X-ray spectra of both stellar-mass and supermassive black holes (see, for instance, Refs.~\cite{Cao:2017kdq,Tripathi:2018lhx,Tripathi:2020dni,Tripathi:2020yts,Zhang:2021ymo}). Such X-ray tests currently provide among the most stringent constraints on possible deviations from the Kerr solution (see, e.g., Fig.~13 in Ref.~\cite{Bambi:2022dtw}). Here we want to extend existing studies to understand modeling uncertainties in tests of the Kerr black hole hypothesis using X-ray reflection spectroscopy.

The impact of some simplifications in current reflection models to test the Kerr hypothesis  has already been discussed in the literature. For example, all the available reflection models assume that the accretion disk is geometrically thin and the motion of the gas in the disk is Keplerian. If we use our reflection models to fit the data of sources with thick disks, we can easily obtain incorrect estimates of some parameters of the system even if the quality of the fit is good~\cite{Riaz:2019bkv,Riaz:2019kat}, which means that current measurements of black holes with high mass accretion rates are not reliable. On the other hand, reflection models seem to be suitable to analyze current data of sources with thin disks, as it has been tested with observations in Refs.~\cite{Abdikamalov:2020oci,Tripathi:2021wap,Jiang:2022sqv} and with simulations of accretion disks generated by GRMHD codes in Ref.~\cite{Shashank:2022xyh}. The impact of the radiation from the plunging region has been investigated in Refs.~\cite{Zhou:2019dfw,Cardenas-Avendano:2020xtw}. If the plunging region is optically thin, we can observer higher order disk images, which can also contributed to the total observed reflection spectrum of the source~\cite{Zhou:2019dfw}. On the contrary, if the plunging region is optically thick, it should also generate a reflection spectrum, even if such a spectrum would not have reflection features because of the high value of the ionization parameter in the low density plunging region~\cite{Cardenas-Avendano:2020xtw} (see also Refs.~\cite{Shashank:2022xyh,Reynolds:2007rx}). In both cases, the impact of the radiation from the plunging region seems to be negligible in current X-ray reflection spectroscopy measurements, especially in the case of fast-rotating black holes with a small plunging region.

In the present work, we want to study the impact of the returning radiation, namely of the radiation emitted by the disk and returning to the disk because of the strong light bending near black holes. The impact of the returning radiation on the estimate of the model parameters in Kerr spacetime was already investigated by two of us in Ref.~\cite{Riaz:2020zqb}, where we concluded that the estimate of most parameters is not significantly affected if the theoretical model to fit the data does not include the calculation of the returning radiation (see also the results in Ref.~\cite{Wilkins:2020lxr}). Here we study the impact of the returning radiation on our tests of the Kerr black hole hypothesis. In our study, we employ the latest version of {\tt relxill} in which the emissivity profile generated by a lamppost corona is calculated including the returning radiation~\cite{Dauser:2022zwc} and we simulate some observations of a bright Galactic black hole with \textsl{NuSTAR}~\cite{NuSTAR:2013yza} (as an example of a current X-ray mission) and \textsl{eXTP}~\cite{eXTP:2016rzs} (as an example of future X-ray mission). As shown in the next sections, we find that our model {\tt relxill\_nk} without returning radiation can recover well the parameters of the spacetime even in the case of future observations with \textsl{eXTP}. We conclude that current constraints on the Kerr metric reported in the literature with {\tt relxill\_nk} are not appreciably affected by the returning radiation.

The paper is organized as follows. In Section~\ref{s-lp}, we briefly review the lamppost coronal model, which is the coronal model assumed in this paper and for which the latest version of {\tt relxill} includes the calculation of the returning radiation. In Section~\ref{s-sim}, we present our simulations and fits. Our results are discussed in Section~\ref{s-dc}.

\section{Returning radiation in the lamppost corona model}\label{s-lp}

\begin{figure*}[th]
    \centering
\includegraphics[trim={0.5cm 0.5cm 0 0cm},clip, width=0.9\textwidth]{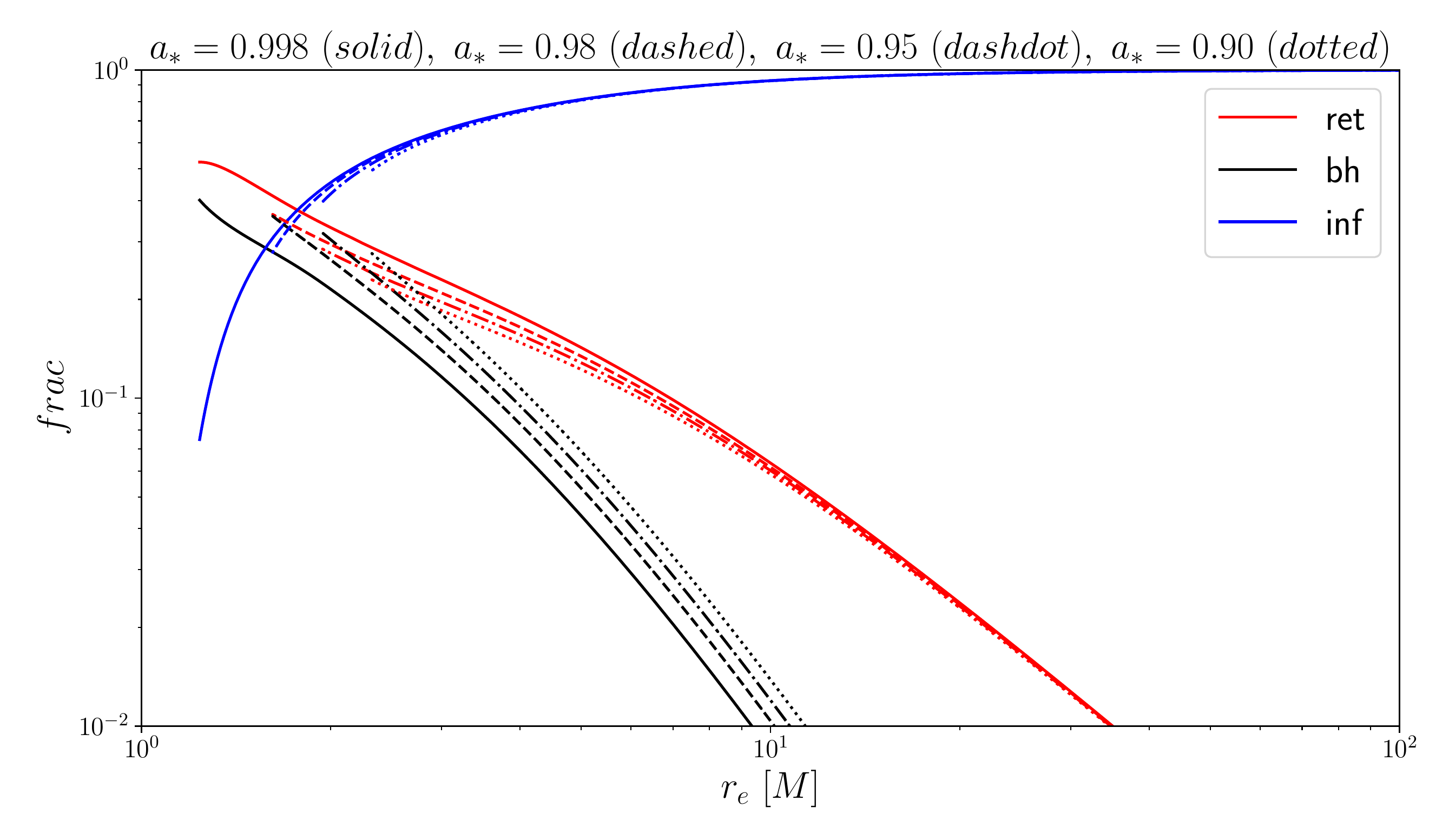}
\caption{Fraction of the radiation returning back to the accretion disk (red lines), falling into the black hole or the plunging region (black lines), and escaping to infinity as a function of emission point on the disk. These fractions are independent of the emissivity profile of the disk. The solid, dashed, dash-dot, and dotted lines represent, respectively, black hole spin $a_* = $ 0.998, 0.98, 0.90, and 0.}
\label{ret-frac}
\end{figure*}

As previously stated, the returning radiation refers to the radiation emitted by the disk that returns back to the accretion disk due to strong light bending. There could be a significant fraction of photons returning back to the disk. We employed our raytracing code to compute the single photon trajectory in the Kerr spacetime to determine the fraction of returning photons. In our raytracing algorithm, we fired photons from a grid of radii, called emission radii $r_{\rm e}$, spanning over the whole accretion disk. The emission of photons in the gas frame is assumed to be isotropic. The fraction of the returning radiation per radial bin $r_{\rm o}$ is then obtained by adding the contribution from all emission points. We incorporate special and general relativistic effects in our calculations. Fig.~\ref{ret-frac} shows the fraction of radiation returning to the disk, falling into the black hole or the plunging region (i.e. the region between the innermost stable circular orbit, or ISCO, and the event horizon), and escaping to infinity. It is important to note that the computation of these fractions does not incorporate the corona's emissivity profile.  The results of Fig.~\ref{ret-frac} are well in agreement with those of Figure 2 in Ref.~\cite{Dauser:2022zwc}. For a maximally rotating black hole,  $a_*$ = 0.998, near the inner edge of the disk approximately 50\% of the emitted photons return back to the disk, up to 40\% of photons either enter the black hole or fall in the plunging region, and only a small fraction of photons escape to infinity. For a given spin, the fraction of the former two radiation components decreases with increasing the emission radius, $r_{\rm e}$, while the fraction of the latter component increases. This is because the effect of gravitational light bending is weaker at larger radii. For a given emission radius and decreasing the black hole spin, the fraction of returning photons decreases, the fraction of photons captured by the black hole increases, and the fraction of photons escaping to infinity remains almost the same.

\begin{figure*}[tb]
    \centering
\includegraphics[width=0.82\textwidth]{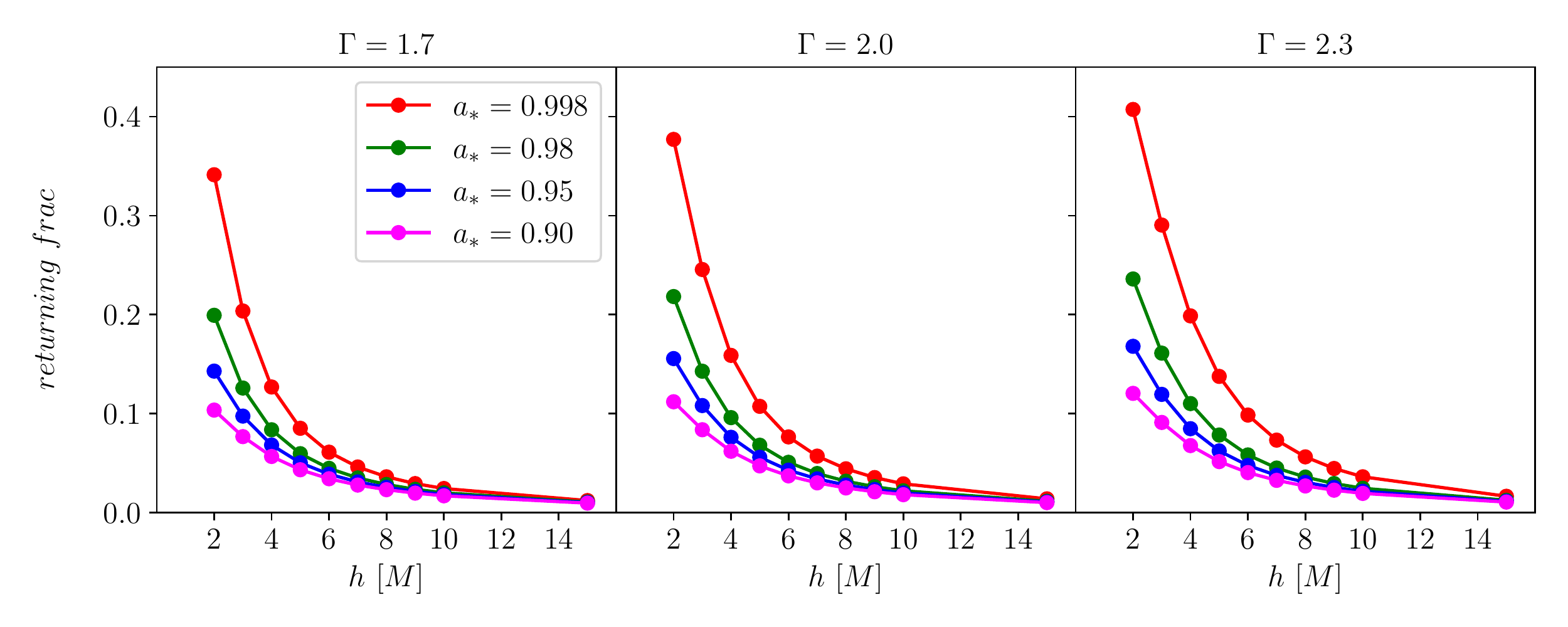}
\caption{Total fraction of radiation returning back to the accretion disk for different values of the coronal height in the lamppost scenario.  The colors red, green, blue, and magenta correspond to black hole spins $a_* = $ 0.998, 0.98, 0.90, and 0, respectively. The illumination profile of the lamppost corona is calculated using a photon index of $\Gamma = $ 1.7 (left panel), 2.0 (middle panel), and 2.3 (right panel). }
\label{tot-ret-frac}
\end{figure*}

The geometry of the corona determines the emissivity profile of the reflection spectrum of the accretion disk\footnote{The emissivity profile is the variation of the reflected bolometric flux with disk radius~\citep{Riaz:2020svt, Dauser:2022zwc}.}. Assuming energy conservation in the reprocessing of radiation in the disk material, the fluxes of reflected and incident radiation are equal~\citep{Dauser:2022zwc}. The geometry of the corona in the vicinity of accreting black holes remains unknown. However, several configurations have been proposed in the literature~\citep{Wilkins:2012zm,Gonzalez:2017gzu,Riaz:2020svt}. When the coronal geometry is unknown, the emissivity profile can be modeled using an empirical radius-dependent power-law or a broken power-law. In the case of a certain coronal geometry, we can calculate the emissivity profile in terms of a few parameters describing the corona. The lamppost model is currently the most popular coronal geometry, in which the corona is supposed to be a point-like source at a specific height $h$ along the black hole spin axis~\citep{Dauser:2013xv, Riaz:2020svt, Niedzwiecki:2016ncz}. This setup naturally produces highly focused irradiation of the inner accretion disk for a source positioned at a low height above the black hole, as it has been found in many observations~\citep{Dauser:2013xv, Riaz:2020svt, Niedzwiecki:2016ncz}. As a consequence, a notable fraction of the radiation returns back to the accretion disk.  Fig.~\ref{tot-ret-frac} shows the total fraction -- over the entire accretion disk of size 1000~$M$ -- of photons that return to the accretion disk as a function of the corona's height for different values of the black hole spin. For a maximally spinning black hole, at the corona's lowest height, the total returning fraction exceeds 30\%, which decreases to an almost insignificant fraction beyond $h = $ 10~$M$. This phenomenon is attributed to the fact that, at an increased height of the corona, the illumination level on the inner portion of the disk is comparatively reduced. Likewise, the returning fraction decreases at a lower spin of the black hole. The reason for this is that when the spin is decreased, the ISCO radius, which designates the location of the disk's inner edge, shifts towards a larger radius. Furthermore, it is noteworthy that the total returning fraction exhibits an increasing trend as the photon index ($\Gamma$) of the corona's continuum increases because the emissivity profile from the corona follows $\varepsilon(r) \propto g^{\Gamma}$~\citep{Riaz:2020svt, Dauser:2022zwc}.

Due to intense gravitational light bending at the inner disk, a fraction of reflected photons returns to the other part of the disk, illuminating the disk and producing the so-called secondary reflection~\citep{Riaz:2020zqb,Dauser:2022zwc}. This secondary reflection inevitably produces a distortion in the total reflection spectrum, which, if ignored in the model computation, may lead to a systematic bias in the final parameter estimations~\citep{Riaz:2020zqb}, and may also affect tests of General Relativity and of the Kerr metric. Recently, the model {\tt relxilllp}, the version of {\tt relxill} with a lamppost corona, has been updated to include the effect of returning radiation~\cite{Dauser:2022zwc}. Within some simplifications to include the returning radiation in the model, the authors of Ref.~\cite{Dauser:2022zwc} find that the effect of returning radiation is greater for a fast-rotating black holes and a lower lamppost height, as it could have been expected. They also find that the main impact of returning radiation is to flatten the emissivity profile.

\begin{table*}[tbh]
\renewcommand{\arraystretch}{1.5}
\centering
\begin{tabular}{c c c |c c |c c c c c}
\hline\hline
\vspace{-0.2cm}
\hspace{0.3cm} $a_*$ \hspace{0.3cm} & \hspace{0.3cm} $h~[M]$ \hspace{0.3cm}& \hspace{0.3cm} $i~[{\rm deg}]$ \hspace{0.3cm}& \multicolumn{2}{c|}{\hspace{0.3cm} X-ray mission(s) \hspace{0.3cm}}&& \multicolumn{2}{c}{\hspace{0.3cm} fit \hspace{0.3cm}} \hspace{0.5cm}\\
& & & & \\ 
\hline
0.98 & 2 &$25, 75$ & \textsl{eXTP}, & \textsl{NuSTAR} && lp, & lp$\_{\rm nk}$  \\
0.98 & 3&$25, 75$ & -- & \textsl{NuSTAR} && lp, & lp$\_{\rm nk}$ \\
0.98 & 4&$25, 75$ & \textsl{eXTP}, & \textsl{NuSTAR} && lp, & lp$\_{\rm nk}$ \\
0.98 & 5&$25, 75$ & -- & \textsl{NuSTAR} && lp, & lp$\_{\rm nk}$ \\
0.98 & 6&$25, 75$ & \textsl{eXTP}, & \textsl{NuSTAR} && lp, & lp$\_{\rm nk}$ \\
0.98 & 7&$25, 75$ & -- & \textsl{NuSTAR} && lp, & lp$\_{\rm nk}$ \\
0.98 & 8&$25, 75$ & \textsl{eXTP}, & \textsl{NuSTAR} && lp, & lp$\_{\rm nk}$ \\
0.98 & 9&$25, 75$ & -- & \textsl{NuSTAR} && lp, & lp$\_{\rm nk}$ \\
0.98 & 10&$25, 75$ & \textsl{eXTP}, & \textsl{NuSTAR} && lp, & lp$\_{\rm nk}$ \\
\hline\hline
\end{tabular}
\vspace{0.2cm}
\caption{Input parameters and X-ray missions used to generate the simulated observations are listed in the first and second columns, respectively. The models used to fit the simulated data are in the final column.   \label{input-params}}
\end{table*}

\section{Simulations and parameter estimates}\label{s-sim}

As demonstrated in the preceding section, the returned radiation may constitute a non-negligible fraction in certain scenarios. However, current publicly available relativistic reflection models, with the exception of {\tt relxilllp} from the {\tt relxill} v2.0 package~\citep{Dauser:2022zwc}, ignore the influence of returning radiation in the calculation of the reflection spectrum. When such models are used to analyze reflection-dominated spectra, they may introduce systematic bias into the system's final parameter estimates~\citep{Riaz:2020zqb}. This can have an impact even in current tests of the Kerr spacetime.

Here we analyze the possible systematic bias in the model's output parameters and tests of General Relativity by simulating observations using theoretical reflection spectra with returning radiation and then fitting them with a model that does not include the returning radiation. To generate the theoretical spectra of accreting black hole-disk system, we employ the following model:
\begin{align*}
{\tt tbabs \times (powerlaw + relxilllp)} , 
\end{align*}
where {\tt tbabs}~\citep{tbabs} takes into account the effect of Galactic absorption along the line of sight, {\tt powerlaw} models the continuum from the corona, and {\tt relxilllp} is the disk's relativistic reflection spectrum, including the effect of returning radiation, detected by the distant observer in a Kerr spacetime.

As a next step, we need to set up the system's configuration by choosing the simulating model's input parameters. For the Galactic absorption, {\tt tbabs} requires a single parameter, the hydrogen column density, $N_{\rm H}$, which is set to $6.74\times10^{20}$ cm$^{-2}$. The power law component needs two parameters: the photon index $\Gamma$, which we set to 1.7, and the high-energy cutoff $E_{\rm cut}$, which we set to 300~keV. The reflection component is generated using the {\tt relxilllp} model with the returning radiation parameter enabled. The spin parameter ($a_{*~\rm input}$) in {\tt relxilllp} is set to 0.98, which is close to the maximum value that amplifies the influence of returning radiation on the total spectrum. The height of the corona ($h_{\rm input}$) is taken from 2 to $10~M$ with a step-size of $1~M$, and the observer's inclination angle ($i$) is set to  $25^{\circ}$ or $75^{\circ}$ (see Tab.~\ref{input-params}). The reflection fraction ($R_{\rm frac~input}$) is fixed at $-1$, so the output is only a reflection spectrum. For the accretion disk’s parameters, the inner ($R_{\rm in}$) and outer ($R_{\rm out}$) edges are set to ISCO and 400 $M$, respectively, the ionization parameter (log$\xi$) is set to 3.1 ($\xi$ in the units of erg cm s$^{-1}$), and the iron abundance ($A_{\rm Fe}$) is set to 1 (Solar abundance).

We assume the observation of a bright Galactic X-ray binary system, and we set the energy flux around $1\times10^{-8}$ erg cm$^{-2}$ s$^{-1}$ in the 1-10 keV energy range. We impose that about half of the photons are from the continuum of the corona and half from the reflection spectrum of the disk. The {\tt fakeit} command in {\tt xspec}~\cite{xspec} is utilized to simulate observations. Regarding X-ray instruments, we consider independent observations with the current mission \textsl{NuSTAR} and the future mission \textsl{eXTP}. The exposure time is set to 40~ks for both missions. This results in about 16~million counts per Focal Plane Module (FPMA and FPMB) for \textsl{NuSTAR} in the energy range of 3-79~keV and approximately 140~million counts for \textsl{eXTP} in the energy range of 2-30~keV.

The simulated observations are analyzed in {\tt xspec} with the models:    
\begin{align*}
&{\tt tbabs \times relxillp}, \\ 
&{\tt tbabs \times relxilllp\_nk}.
\end{align*}
{\tt relxilllp} is still the lamppost corona model of {\tt relxill}, but now the returning radiation is switched off. {\tt relxilllp\_nk} is the lamppost corona model of {\tt relxill\_nk}~\citep{Abdikamalov:2019yrr}, where the spacetime is allowed to have deformations from the Kerr solution and the returning radiation is not included. Here we use the version of {\tt relxilllp\_nk} in which the spacetime geometry is described by the Johannsen metric with only one possible non-vanishing deformation parameter, $\alpha_{13}$~\citep{Johannsen:2013szh}. In the fitting procedure, $N_{\rm H}$, $R_{\rm in}$, $R_{\rm out}$, and $E_{\rm cut}$ are fixed at their input values, while all other parameters are permitted to vary freely.

\begin{figure*}[th]
    \centering
\includegraphics[width=0.80\textwidth]{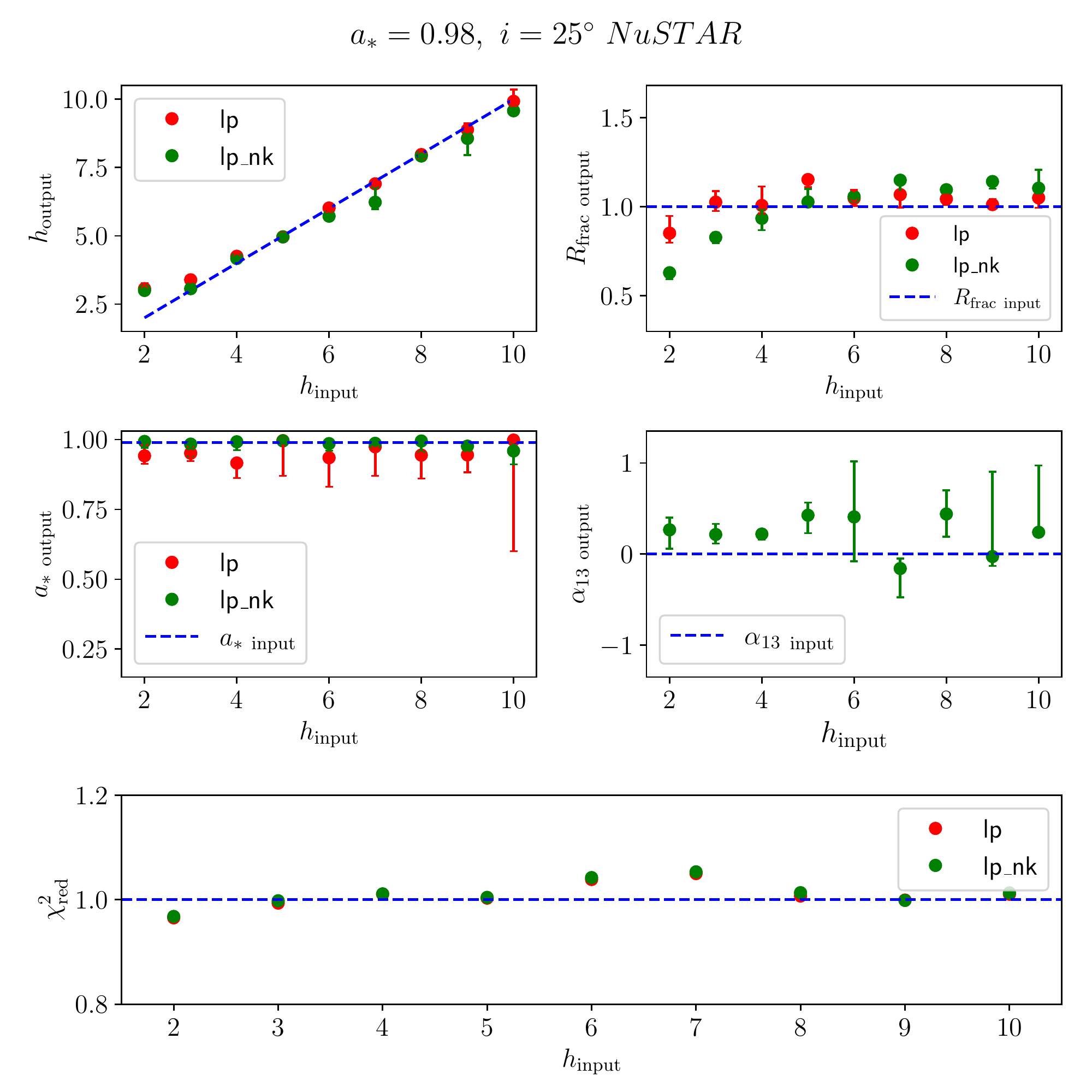}
\caption{Best-fit value of lamppost height ($h_{\rm output}$, upper left panel), reflection fraction ($R_{\rm frac}$, upper right panel), black hole spin ($a_*$, middle left panel), deformation parameter ($\alpha_{13}$, middle right panel), and the reduced chi-square ($\chi^2_{\rm red}$, bottom panel) for simulated observations with \textsl{NuSTAR} for an observer's inclination angle $i = 25^{\circ}$. The colors red and green represent the employed model $\tt{relxilllp}$ and $\tt{ relxilllp\_nk}$, respectively, and they do not include the returning radiation in the fit.  The input spin parameter in the simulations is fixed at $a_* = 0.98$. The error bar on each parameter corresponds to the statistical uncertainty of 90\% confidence level. If there is no error bar, it is either too small at this scale or cannot be calculated. }
\label{param-NS-25}
\end{figure*}

\begin{figure*}[th]
    \centering
\includegraphics[width=0.80\textwidth]{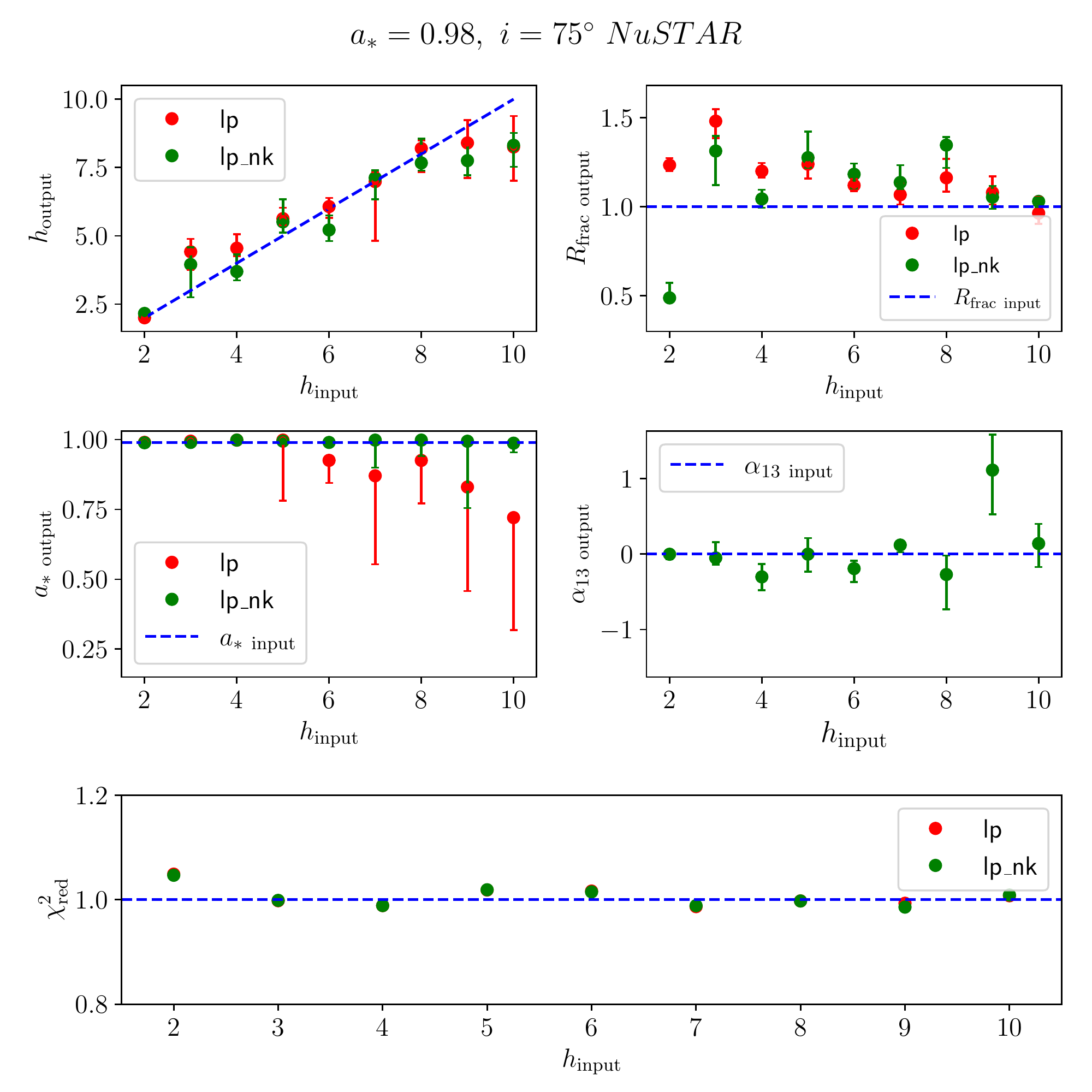}
\caption{As in Fig.~\ref{param-NS-25} but with an inclination angle $i = 75^{\circ}$.  \label{param-NS-75}}
\end{figure*}

\begin{figure*}[tbh]
    \centering
\includegraphics[width=0.80\textwidth]{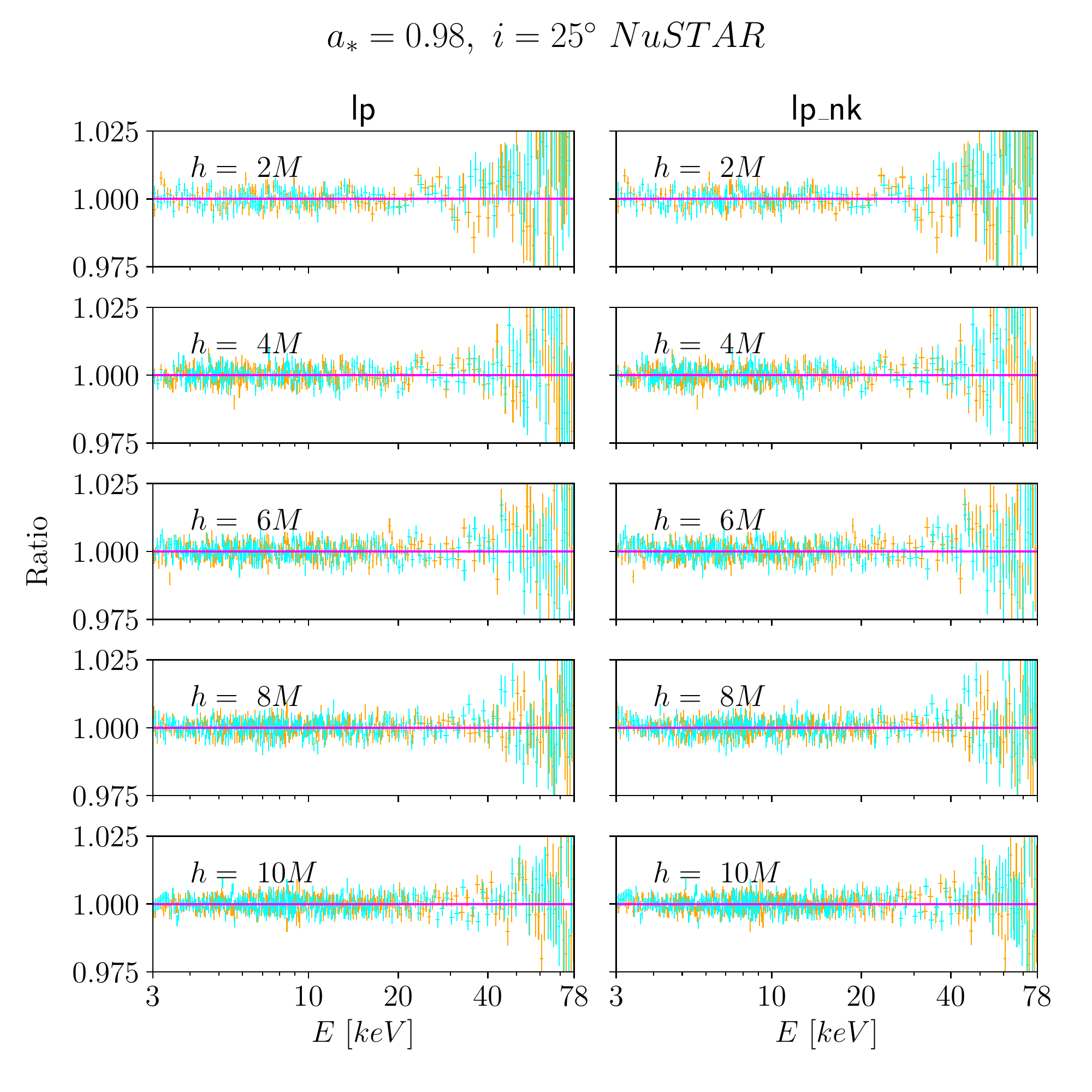}
    \caption{Ratio between data and best-fit model for simulated observations with \textsl{NuSTAR} with viewing angle $i = 25^{\circ}$. \label{ratio-25-NuSTAR}}
\end{figure*}

\begin{figure*}[tbh]
    \centering
\includegraphics[width=0.80\textwidth]{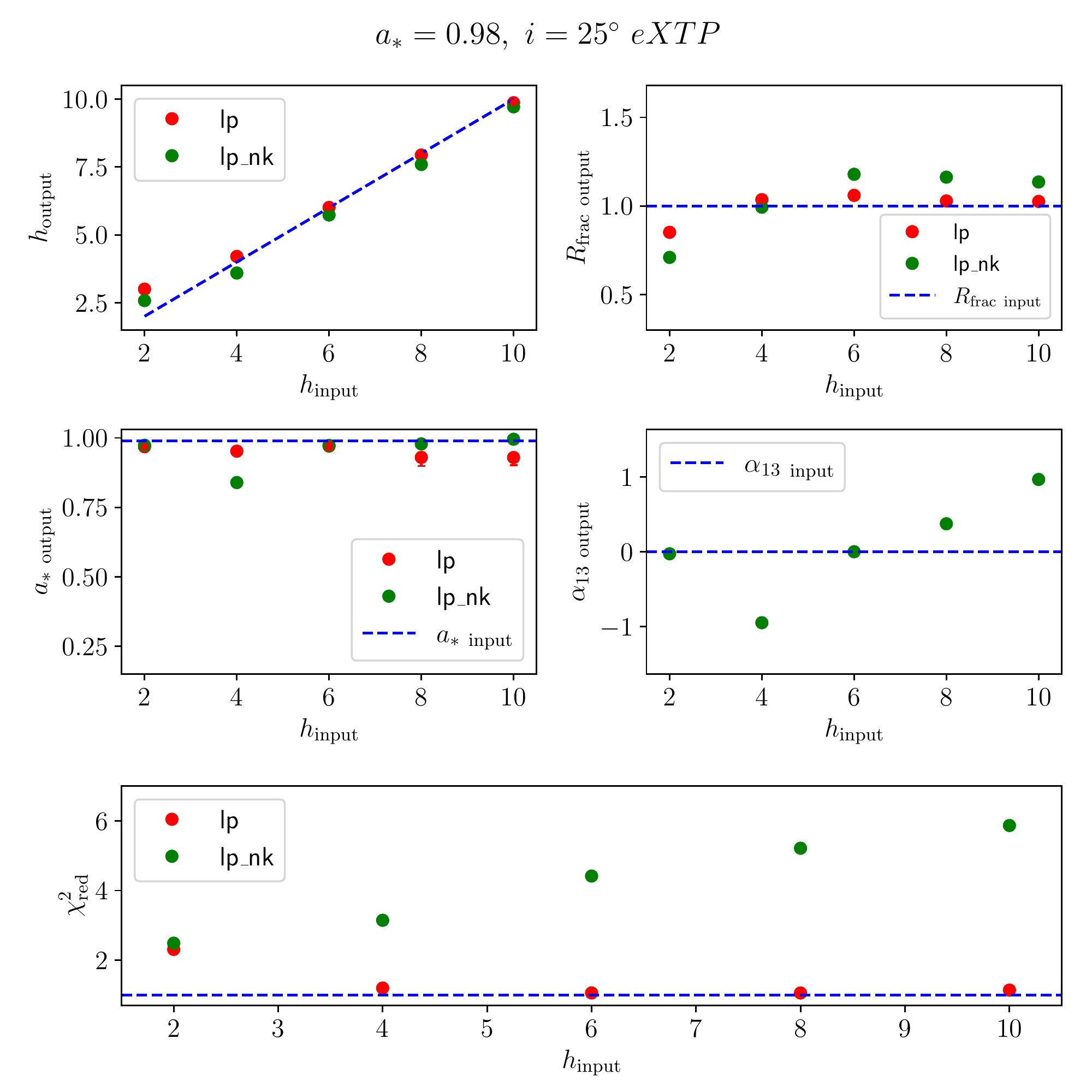}
\caption{As in Fig.~\ref{param-NS-25} but for the \textsl{eXTP} simulations with inclination angle $i = 25^{\circ}$. \label{param-eXTP-25}}
\end{figure*}

\begin{figure*}[tbh]
    \centering
\includegraphics[width=0.80\textwidth]{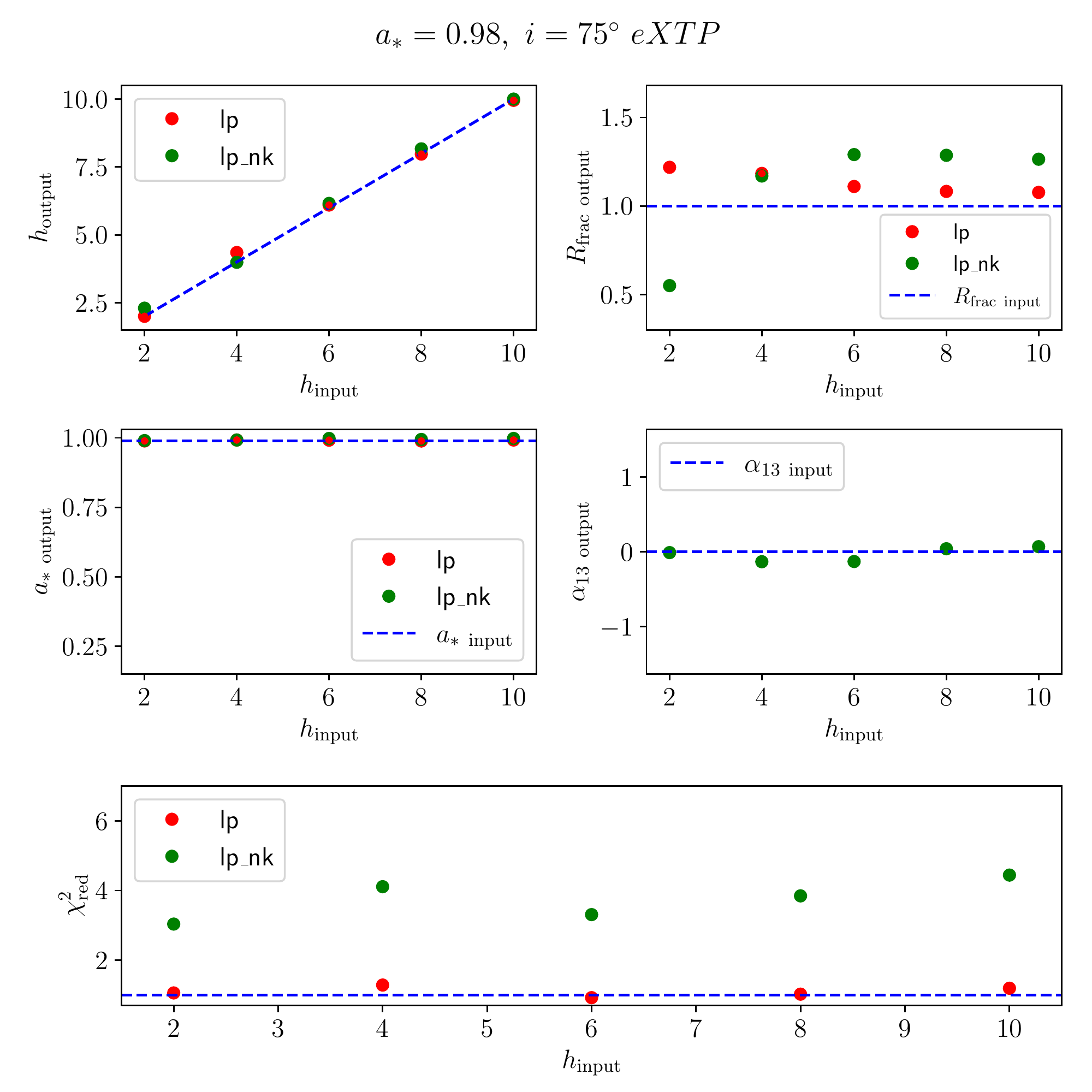}
\caption{As in Fig.~\ref{param-NS-25} but for simulated observations with \textsl{eXTP} with inclination angle $i = 75^{\circ}$. \label{param-eXTP-75}}
\end{figure*}

\begin{figure*}[tbh]
    \centering
\includegraphics[width=0.80\textwidth]{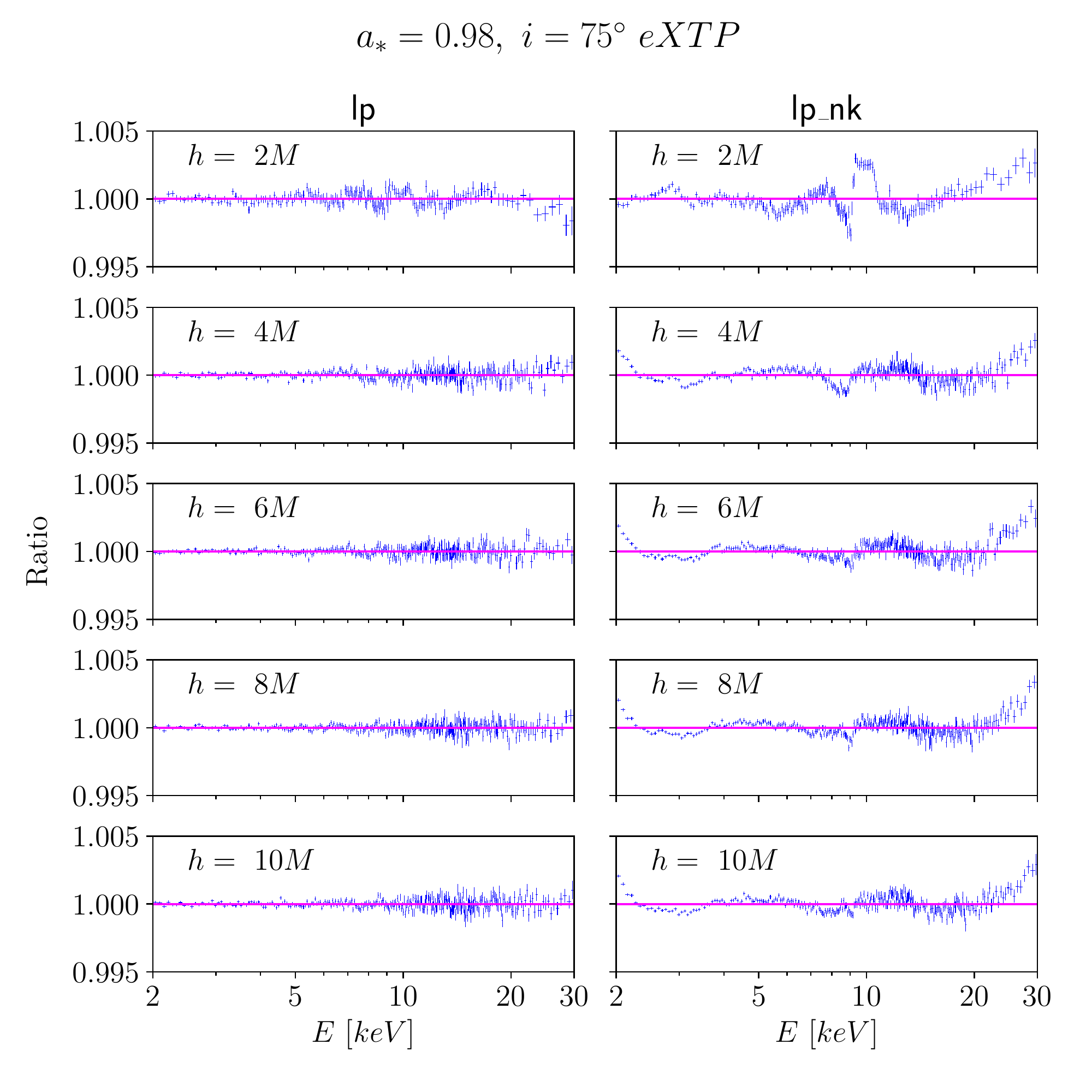}
    \caption{Ratio between data and best-fit model for simulated observations with the future X-ray mission \textsl{eXTP} for a disk's inclination angle $i = 25^{\circ}$. \label{ratio-25-extp}}
\end{figure*}

\begin{figure*}[tbh]
    \centering
\includegraphics[width=0.80\textwidth]{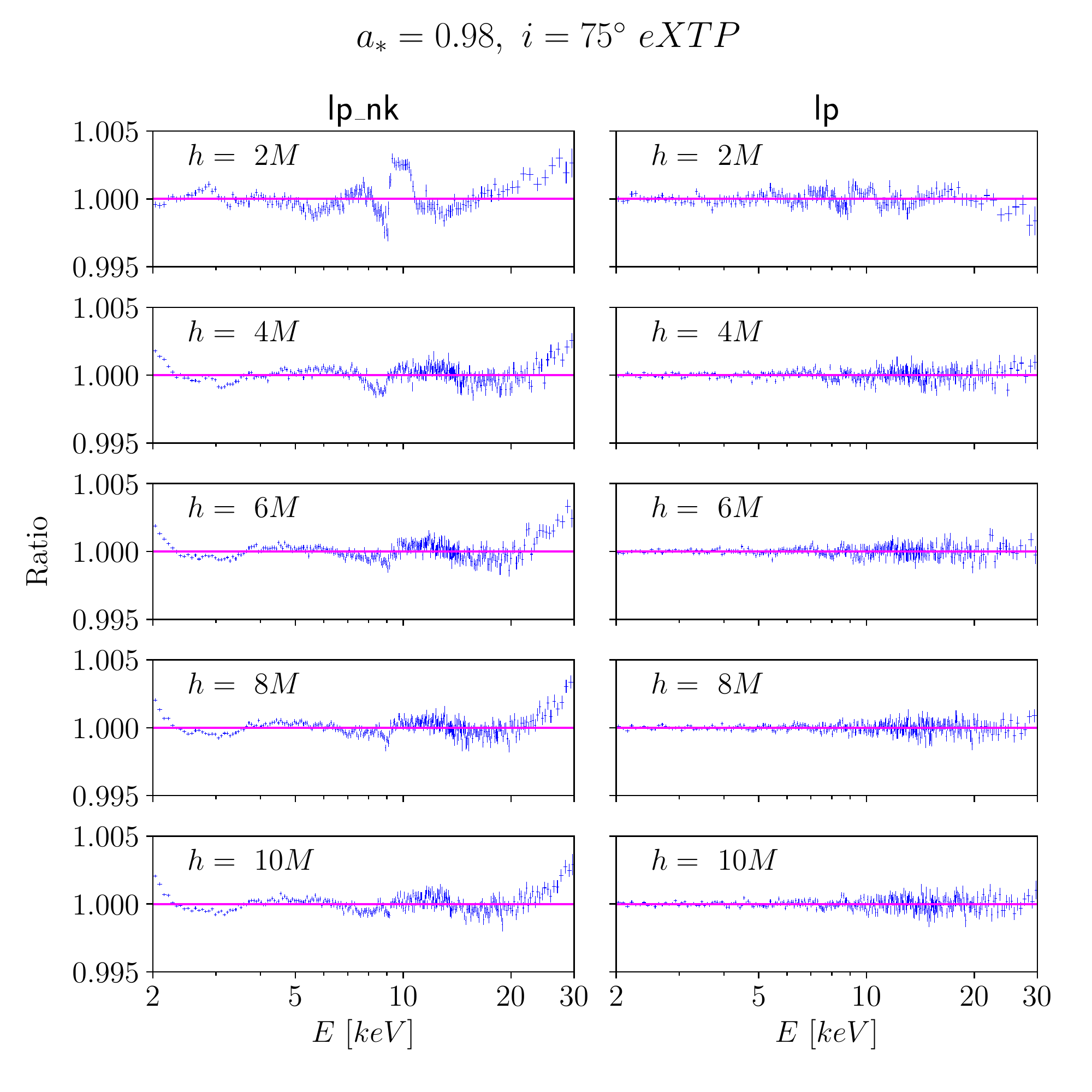}
    \caption{As in Fig.~\ref{ratio-25-extp} but in the case of a disk's inclination angle $i = 75^{\circ}$. \label{ratio-75-extp}}
\end{figure*}

The ratios between data and the best-fit models are shown in Fig.~\ref{ratio-25-NuSTAR} for the \textsl{NuSTAR} simulations and in Figs.~\ref{ratio-25-extp} and~\ref{ratio-75-extp} for the \textsl{eXTP} ones. The best-fit values of the parameters are shown in Figs.~\ref{param-NS-25} and~\ref{param-NS-75} for the case of \textsl{NuSTAR} and in Figs.~\ref{param-eXTP-25} and~\ref{param-eXTP-75} for the case of \textsl{eXTP}. If present, the error bars associated with the measurements of the parameters correspond to the statistical uncertainty of 90\% confidence level. If absent, they are either too small at this scale or cannot be calculated. The discussion of these results is postponed to the next section.

\section{Discussion and Conclusion}\label{s-dc}

Let us first discuss the results of the \textsl{NusTAR} simulations. The quality of the fits turns out to be good for both values of the inclination angle: there are no unresolved features in the ratio plots (see Fig.~\ref{ratio-25-NuSTAR}) and the reduced chi-square ($\chi^{2}_{\rm red}$) is close to 1 (see the bottom panel in Figs.~\ref{param-NS-25} and~\ref{param-NS-75}). Therefore, we omit to show the ratio plots of the high inclination angle simulations. Overall, most of the input parameters in the simulations are recovered well within the 90\% confidence limit. The deformation parameter ($\alpha_{13}$) is close to zero in most of the simulations, indicating that we recover the Kerr metric (which is the metric assumed in the simulations). We should also note that these simulated observations are intended to be more optimistic than actual observations of accreting black holes and that a small deviation from the Kerr metric may be caused by the response of the instrument or a statistical fluctuation. The impact of returning radiation on the reflection fraction parameter is more obvious. For the lowest corona height, it is overestimated (underestimated) at a high (low) inclination angle. This is because, at a low height, the contribution of the reflection spectrum due to returning radiation is higher, producing a distortion in the spectrum, which is compensated by adjusting the reflection fraction parameter. However, as the lamppost height increases, the reflection fraction approaches its input value. This is because the inner part of the accretion disk is relatively less illuminated at higher heights, which results in lesser primary reflection photons in the strong gravity region and lesser being returned to the accretion disk.

For the \textsl{eXTP} simulations, we discuss separately the fits with {\tt relxilllp} (Kerr metric) and those with {\tt relxilllp\_nk} (when we allow for deviations from the Kerr metric). When the model of the fit assumes the Kerr metric, the quality of the fit is good, $\chi^{2}_{\rm red}$ is close to 1 (see the bottom panel in Figs.~\ref{param-eXTP-25} and~\ref{param-eXTP-75}), and there are no unresolved features in the ratio plots (see Figs.~\ref{ratio-25-extp} and~\ref{ratio-75-extp}). The simulations with $h = 2~M$ are an exception and we will discuss them later. In general, the fit successfully recovers the input parameters. However, the reflection fraction exhibits a similar trend to that of the \textsl{NuSTAR} simulations. At the lowest height, it is underestimated for the low inclination angle and overestimated for the high inclination angle. In contrast to the \textsl{NuSTAR} case, the lowest coronal height simulations now show some unresolved features in ratio plots. This is due to the combined effect of the greatest influence of returning radiation at the lowest coronal height and improved data quality, which can no longer be compensated by adjusting the reflection fraction parameter.

Considering the {\tt relxilllp\_nk} fits, the quality of the fit is poor; the ratio plot shows unresolved features, and $\chi^{2}_{\rm red}$ is greater than 1. We recover the Kerr metric for simulations with a high inclination angle. At a low inclination, the Kerr metric is not recovered in the majority of cases. However, such a low quality of the fits and the incapability of recovering the Kerr metric for low viewing angles seem to be due to a small discrepancy between the lamppost emissivity profiles of {\tt relxilllp} and {\tt relxillp\_nk}, which shows up only for high-quality data. Otherwise, $\chi^2$ of {\tt relxillp\_nk} could not be higher than the $\chi^2$ of {\tt relxillp} because we have the same model and one more free parameter. The lamppost emissivity profiles of {\tt relxilllp} and {\tt relxillp\_nk} were indeed compared in \cite{Abdikamalov:2019yrr} for the quality of current X-ray missions, finding consistent predictions. However, we see that some disagreement shows up when we consider higher quality data.

In conclusion, the study presented in this paper suggests that current constraints on the Kerr metric with the reflection model {\tt relxillp\_nk} are not appreciably affected by the returning radiation, which is not included in the calculations of the model. Such a result was not obvious because the returning radiation can count for more than 10\% in many realistic cases (e.g., $a_* > 0.95$ and $h < 4 \, M$) and up to more than 30\% in some extreme cases (e.g., $a_* = 0.998$ and $h = 2 \, M$). We note that our simulations and fitting model assume the popular lamppost corona, in which the corona is a point-like source along the black hole spin axis. This is certainly another simplification of the model because the corona must have a finite size in order to be able to Comptonize a sufficiently high fraction of thermal photons from the disk. Moreover, some systems certainly do not have a compact corona along the black hole spin axis (see, for instance, the recent polarization results in Ref.~\cite{Krawczynski:2022obw}). The coronal geometry is certainly another important ingredient of current reflection models and can have a strong impact on the accuracy of the final measurements of the parameters of the system. However, we leave the study of the systematic uncertainties induced by the assumptions on the coronal geometry to future studies.

\vspace{0.5cm}

{\bf Acknowledgments --}
This work was supported by the Shanghai Municipal Education Commission, Grant No. 2019-01-07-00-07-E00035, the Natural Science Foundation of Shanghai, Grant No. 22ZR1403400, the National Natural Science Foundation of China (NSFC), Grant No.~12250610185, 11973019, and 12261131497, and Fudan University, Grant No. JIH1512604. S.R. acknowledges the support from the China Postdoctoral Science Foundation, Grant No. 2022M720035.

\end{document}